\documentstyle[aps,epsf,prb]{revtex}

\def\beq{\begin{equation}}
\def\eeq{\end{equation}}

\begin{document}

\title{Quantum Glass Transition in a Periodic Long-Range Josephson Array}

\author{D. M. Kagan$^{1,2}$, L. B. Ioffe$^{1,3}$ and M. V. Feigel'man$^1$}
\address{$^1$Landau Institute for Theoretical Physics, Moscow, 117940, RUSSIA}
\address{$^2$ABBYY, Moscow, p.b.\#19, 105568, RUSSIA}
\address{$^3$Department of Physics, Rutgers University, Piscataway, NJ 08855, USA}
\date{\today}
\maketitle

\begin{abstract}
We show that the ground state of the periodic long range Josephson array frustrated
by magnetic field is a glass for a sufficiently large Josephson energies despite the
absence of a quenched disorder. Like superconductors, this glass state has non-zero
phase stiffness and Meissner response; for smaller Josephson energies the glass "melts"
and the ground state loses the phase stiffness and becomes insulating. We find the
critical scaling behavior near this quantum phase transition: the excitation gap vanishes
as $(J-J_c)^2$, the frequency-dependent magnetic susceptibility behaves as
$\chi(\omega) \propto \sqrt{\omega}\ln{\omega}$.
\end{abstract}

\section{Introduction}

Glass formation in the absence of {\it intrinsic} disorder
is a long-standing problem but the last years saw a rapid progress
\cite{Parisi1,Parisi2,Bouchaud,Chandra95,Chandra96a,Chandra96b,Chandra97} in the
qualitative understanding of this phenomena.
Mostly this progress is due to the solutions of periodic models that
assume a mapping between the periodic model and the appropriate random model
\cite{Parisi1,Parisi2,Bouchaud}. The validity of this assumption is still an open
question in a general case but it was shown that at least one periodic model
allows a direct study of a phase transition \cite{Chandra96a} and non-ergodic
behavior below the transition \cite{Chandra97} without any reference to a disordered
model. This model describes a long-range Josephson array in a magnetic field and another
reason for the interest in this model is that it  that can be realized
experimentally (cf. \cite{Chandra96b,Tinkham} for the discussion of experimental
conditions).

All these results were obtained in the framework of classical statistical mechanics, the
glass formation in a {\it regular quantum} systems has not been addressed. The goal of
this paper is to fill this gap. The problem of a glass formation in {\em disordered}
quantum systems was discussed in a number of works  \cite{Huse,SachRead,SachRead2};
these works studied the critical behavior near the quantum vitrification transition
\cite{Huse,SachRead} and the properties of the glassy phase itself \cite{SachRead2}
using the replica approach. They found that the glass  phase transition at
$T=0$  indeed exists; further, it strongly resembles classical (high $T$) phase
transition in the same system: the main difference is in the critical
exponent of correlation function which decays faster than at the classical
critical point:
$D(t)= \overline{\langle S_j(0)S_j(t)\rangle}
 \sim t^{-1}$ at $T=0$ (cf. $D(t)\sim t^{-1/2}$ at non-zero $T$).
A surprising result stated in \cite{SachRead2} is that
at zero temperature no replica-symmetry-breaking (RSB) is needed for the
description of the glassy state, i.e.  replica-symmetrical solution is
stable at $T=0$.  Since, usually, RSB is believed to be a signature of non-ergodicity,
this result means either absence of non-ergodic behavior at $T=0$ or violation of the
usual relation between RSB and non-ergodicity. We feel that in order
to clarify this important question, an approach that is free from the ambiguities of
the replica method should be employed.

Understanding of a quantum glass formation in a system with regular Hamiltonian
is important for the general problem of Quantum Computation \cite{quantcomp}. The reason
is that quantum computer is also a quantum system with exponential number of states and the
process of computation can be viewed as an almost adiabatic change of the external parameters
resulting in a different state. The crucial question is what are the conditions so that such
process does not lead to the collapse of the density matrix due to the coupling to the
environment. This question can be addressed to the spin glass system as well and one can learn
about decoherence in a generic large system with exponential number of states from the answer
to it.

Here we study the quantum version the long-range Josephson array in a frustrating
magnetic field that was suggested in \cite{Chandra95,Chandra96a,Chandra96b,Chandra97}.
We consider here only the problem of glass formation, approaching the glass from the
"liquid" (i.e. insulating) side. We show that the quantum version of this problem is
described by the same dynamic equations as the quantum disordered p-spin model studied
in \cite{Cu98}. Thus, we explicitly prove that this frustrated quantum system can be
mapped onto the quantum disordered system in a complete analogy with the situation for
classical problems. Further, we provide a direct numerical proof that the
transition in this model is indeed continuous as conjectured in \cite{Cu98} and
we calculate the anomaly of the diamagnetic response associated with this
transition.

Another, and more physical, justification of the model is the following.
It is well established, both experimentally (cf. e.g. \cite{delft,chalmers})
and theoretically \cite{JJtheor} that usual nearest-neighbors Josephson arrays made of
small superconductive islands demonstrate zero-$T$
superconductor-insulator transition as the ratio
of the Josephson coupling $E_J$ between the  superconductive islands
to the Coulomb energy cost $E_C = (2e)^2/2C$ for the transfer of the
Cooper pair between the islands decreases. At small values of $x = E_J/E_C$
the ground state is an insulator with nonzero Coulomb gap in the excitation
spectrum. At nearly-critical values of $x \approx x_{cr}$ the transition
between insulating and superconductive states
can be triggered by application of a weak magnetic field, producing
frustration of the Josepson interaction; moreover, this transition can
be splited \cite{delft} into the sequence of two different transitions:
superconductor-metal-insulator.  Although main qualitative features
of these phenomena are understood, there is still no quantitative
theory which describes quantum phase transitions in 2-dimensional
short-range systems, especially in the presence of frustration.
Therefore, in our attempt to study the origin of a quantum glass state,
we have to turn to the simplest (theoretically) model of a Josephson array
with long-range interaction, which
consists of long superconducting wires (instead of islands).
It will allow us to employ some version of mean-field-theory and reduce
the problem to a zero-dimensional quantum theory  with the interaction that is
non-local in  time.

The system of our study is a stack of two mutually perpendicular sets of
$N$ parallel thin superconducting wires with Josephson junctions at each node
that is placed in an external transverse field $H$.
Macroscopic quantum variables
of this array are the $2N$ superconducting phases associated with
each wire (e.g. the value of the phase of the superconducting order parameter at the
center of each wire); we will always
assume that excitations within individual wire can be neglected, so the whole wire
is characterized by one phase, $\phi_m$.
In the absence of an external field the phase differences
would be zero at each junction, but this is not possible for finite
$H$, so the phases are frustrated.  Here we assume that the
Josephson currents are sufficiently small so that the induced fields
are negligible in comparison with $H$ (this imposes an important constraint for the
experimental realization of this network \cite{Chandra96b}).
Therefore the array is described by the Hamiltonian
\beq
{\cal H} = {\cal H}_J + {\cal H}_C =
-E_J\sum_{m,n} \cos(\phi_n-\phi_m - \frac{2e}{\hbar c}\int{\bf A}{d\bf
l}) \quad + \quad \frac{(2e)^2}{2}\sum_{m,n}
 \hat{C}_{m,n}^{-1} \frac{\partial}{\partial \phi_m}
\frac{\partial}{\partial \phi_n}
\label{1}
\eeq
where $H_J$ and $H_C$ represent, correspondingly, Josephson and Coulomb
parts of the Hamiltonian, and $\hat{C}_{m,n}$ is the matrix of the capacitances.
There are several different contributions to $\hat{C}$:
self-capacitances of the wires $C_l$ (with respect to substrate),
the contact capacitances $C_J$ and mutual capacitances of wires
$C_{ll}$. Below we will assume, that the self-capacitance is the
largest of all, $C_l \gg C_{ll}, NC_J$ (the factor $N$ accounts for the
fact that there are $N$ contacts along each wire).
These conditions allow as to neglect all mutual
capacitances and consider the matrix $C_{m,n}$ to be diagonal with
eigenvalues $C_l$.  It is convenient to rewrite the Hamiltonian
in terms of  ``spin'' variables $s_m = e^{i\phi_m}$. Choosing the Landau gauge
for the vector potential and introducing $J_0$ by $E_J = \frac{J_0}{\sqrt{N}}$
so that the transition temperature remains constant in the limit  $N\to\infty$
at fixed $J_0$ we get
\beq
{\cal H} = - \sum_{m,n}^{2N} s_m^{*} {\cal J}_{mn} s_n +
\frac{E_C}{2}\sum_n Q_n^2
\label{H}
\eeq
where $Q_n \equiv -i\partial/\partial\phi_n$ is the charge operator
conjugated to the phase $\phi_n$, $E_C = \frac{4e^2}{C_l}$,
and ${\cal J}_{mn}$ is the coupling matrix
\beq
\hat{\cal  J} = \left( \begin{array}{cc}
0 & \hat{J} \\
\hat{J}^\dagger & 0
\end{array}
\right)
\label{J}
\eeq
with $J_{jk} = \frac{J_0}{\sqrt{N}} \exp(2\pi i \alpha jk /N)$ and $1 \! \leq
\! (j,k) \! \leq \! N$ where $j(k)$ is the index of the horizontal (vertical)
wires;
$s_m = e^{i\phi_m}$  where the $\phi_m$ are the superconducting phases of the
$2N$ wires.
Here $\alpha = NHl^2/\Phi_0$ is  the flux per unit strip,
$l$ is the inter-node spacing and $\Phi_0$ is the flux quantum.

Because every horizontal (vertical) wire is linked to every vertical
(horizontal) wire, the connectivity in this model is high ($N$) and
it is accessible to a mean-field treatment
(its classical version was developed in \cite{Vinokur87,Chandra96a}).
For ${1\over N} \ll \alpha < 1$ there exists an extensive number of metastable solutions
which minimize the Josephson (``potential'') part of the Hamiltonian (\ref{H}); these
minima are separated by the barriers that scale\cite{Chandra95}  with $N$.
A similar (classical) long-range network with disorder was previously found to display a
spin glass transition\cite{Vinokur87} for $\alpha \gg {1\over N}$; in the absence of
short-range phase coherence between wires $(\alpha \gg 1$) it is equivalent to the
Sherrington-Kirkpatrick model.\cite{Sherrington75}
Physically this glassy behavior occurs because the phase differences associated with the
couplings, $J_{jk}$, acquire random values and fill the interval $(0,2\pi)$ uniformly.
For the periodic case, this condition is satisfied in the ``incommensurate
window'' ${1\over N} \ll \alpha \le 1$ for which the magnetic unit cell
is larger than the system size so that the simple ``crystalline'' phase is
inaccessible.\cite{Chandra95}
There are thus no special field values for which the
number of minima of the potential energy are not extensive, in contrast
to the situation for $\alpha  > 1$.
Below we will consider the case $1/N \ll \alpha \ll 1$ only.
As follows from the previous studies
\cite{Chandra95,Chandra96a,Chandra96b,Chandra97}, the characteristic
energy scale related to the potential energy ${\cal H}_J$ is of the order
of the glass transition temperature of the classical system,
$T_G \approx J_0/\sqrt{\alpha}$.  The zero-$T$ transition we study
here is driven by the competition between Josephson and Coulomb energies,
the scale of the latter being $E_C = 4e^2/C_l$.
Thus, we expect that the quantum transition occurs at
$J_0/\sqrt{\alpha} \sim E_C$. Our goal is to show that
such a (continuous) phase transition indeed occurs and to study the
critical behavior near the transition point. Below, in the main part of the
paper, we measure all energies in units of $E_C$, and return
to the physical units only in the final expression for the critical behavior
of the ac diamagnetic susceptibility.

\section{Quantum Locator Expansion}

We are going to develop a diagram technique for the Hamitlonian
(\ref{H}) which will be very similar to the one employed previously
\cite{Chandra96a} for the classical Langevin dynamics of
the same array. The idea is to treat Coulomb part of the Hamiltonian
as the zero-level approximation, and construct expansion in powers of
Josephson coupling constant $J_0$, keeping all the terms of the lowest
order in the coordination number $1/N$. Thus our approach can be considered
as quantum version of the Thouless-Anderson-Palmer~\cite{Thouless77} method.

The diagram technique for the Matsubara Green function
\begin{equation}
G_{m,n}(\tau) = -\langle T_\tau s_m(\tau)s_n^\dagger(0)\rangle
,\ s(\tau) = e^{-\tau {\cal H}}se^{\tau {\cal H}}
\end{equation}
is closely related to the one developed in~\cite{Chandra96a}.
Dyson equation for the frequency-dependent matrix Green function reads
(note that in our units $E_C=1$):
\beq
\bf{G}_\omega =
\frac{1} {\tilde{G}^{-1}_\omega - (\bf{J J^{\dagger}})\tilde{G}_\omega}
\label{Gg}
\eeq
where we introduced the local Green
functions $\tilde{G}_{\omega}$ that is irreducible with
respect to the $J_{ij}$ lines.
The matrix $(J J^\dagger)_{ij}$ depends  only on the ``distance'' $i-j$
and acquires a simple form in Fourier space $(J J^\dagger)_p =
(J_0^2/\alpha) \theta(\alpha \pi - |p|)$; therefore in this representation
\beq
G_{\omega}(p) =  \frac{\theta(\alpha \pi - |p|)}
{\tilde{G}^{-1}_\omega -
        \frac{J_0^2}{\alpha} \tilde{G}_\omega} +
\frac{\theta(|p| - \alpha \pi)}{\tilde{G}_\omega^{-1}}
\label{GG}
\eeq
Diagrammatically Eq.(\ref{Gg}) and the equation for the
irreducible function $\tilde{G}_{\omega}$ are represented by the graphs shown in Fig. 1.

\centerline{\epsfxsize=8cm \epsfbox{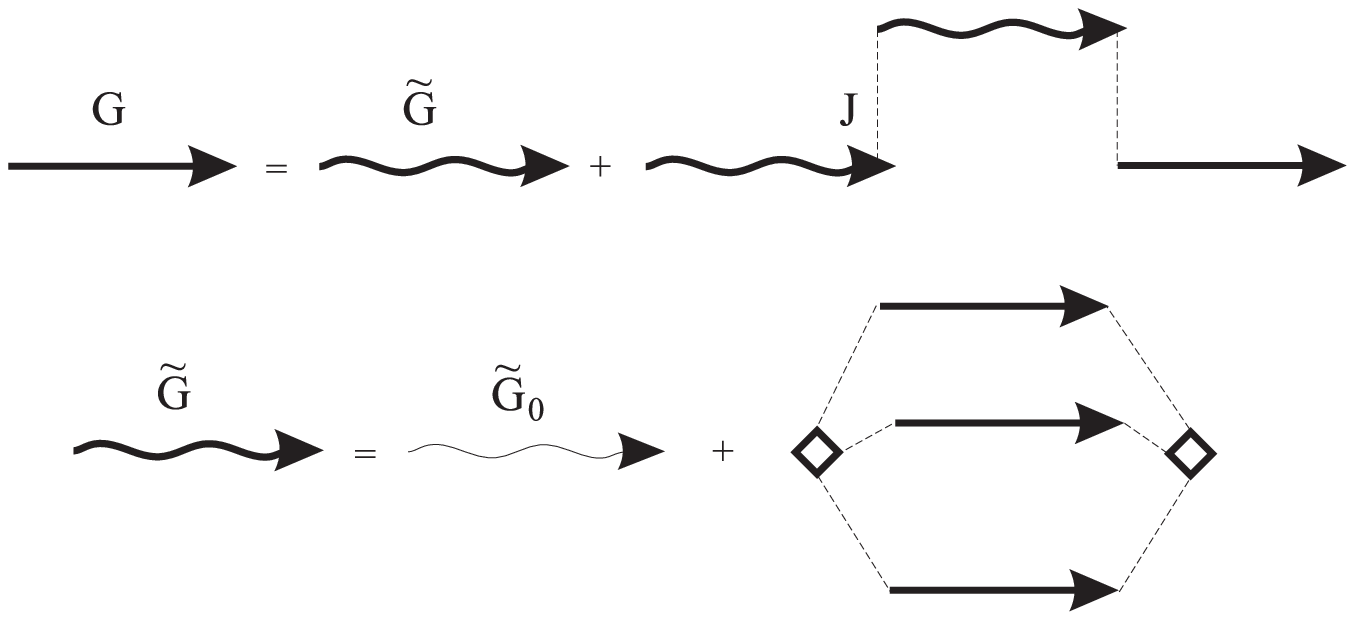}}

\noindent
Note that the equation for $\tilde{G}$ is written in the lowest nontrivial
order in $\alpha$. Indeed, it is seen from Eq.(\ref{GG}) that nontrivial
part of the Green function which contains critical slowing down,
is of relatively small weight $\sim\alpha$. It is this long-time part of
$G_{\omega}$ which enters 3-line diagram on Fig.1 and makes it proportional
to $\alpha^3$; more complicated diagrams either contain even higher powers
of $\alpha$, or are small as $1/N$. Since the second diagram on Fig.1
contains single-site functions only, the whole system of equations
can be written in the form
\begin{eqnarray}
G(\omega) = (1-\alpha)\widetilde G(\omega) + \hat G(\omega); \qquad
\hat G(\omega) = \frac{\alpha \widetilde G(\omega)}{1-J_0^2\widetilde
G^2(\omega)/\alpha}
\label{G_full} \\
\widetilde G(\omega) = \widetilde G_0(\omega) + \Sigma(\omega); \qquad
\Sigma(\omega) = \left(\frac{J_0^2}{\alpha}\right)^3\chi_3^2
\int\limits \hat G^3(t) \exp(i\omega t)\ dt
\label{Sigma}
\end{eqnarray}
Here $\chi_3\sim 1$, as in~\cite{Chandra96a}, is a static value of
four-point vertex denoted as a square box in Fig.1 (we assume that,
 like in~\cite{Chandra96a}, the  main critical  anomaly is contained
in the 2-point Green function alone).
 Equations (\ref{G_full},\ref{Sigma})
 should be solved with obvious initial condition:
\begin{equation}\label{init_cond}
G(t=0) = \int\frac{d\omega}{2\pi} G(\omega) = 1
\end{equation}
Similar normalization condition in the classical problem
was sufficient to determine $\tilde{G}(\omega=0)$ exactly~\cite{Chandra95}.
The same calculation seems to be difficult for the  present quantum problem
 and we will not perform it here. Instead, we will use general properties
of the function $\tilde{G}_0(\omega)$, namely:
i) $\tilde{G}_0(0) \sim 1$, and ii) $\tilde{G}_0(\omega)$ is analitic
at low $\omega$ and has characteristic frequency scale of the order of 1.
In doing so, we will not determine exact position of the phase transition
(i.e. critical value $J_c$ of the coupling strength $J_0$), but we
will show the existence of
continuous transition and find the form of critical scaling.

Let us first analyze equations (\ref{G_full},\ref{Sigma},\ref{init_cond})
omitting the term with $\Sigma$, and using simplest interpolation
$\tilde{G}_0(\omega) = (\lambda + \omega^2)^{-1}$.
Then initial condition~(\ref{init_cond}) gives us the equation for $\lambda$:
\begin{equation}\label{eqfora}
1 = \frac{1}{2\sqrt{\lambda}} + \frac{\alpha}{4\sqrt{%
\lambda - g}}
\end{equation}
where $g = J_0/\sqrt{\alpha}$.
Thus $\lambda \sim 1$ as long as $g \leq 1$. On the other hand, at $g\gg 1$
the solution is $\lambda - g \equiv a \approx (\alpha/4)^2$. The value of
$a$ determines the asymptotic decay rate of the Green function
\beq
G(t) = \displaystyle\frac{\alpha}{\sqrt a}\exp(-|t|\sqrt a)
\label{G0}
\eeq
with $\Sigma$ being neglected.  It will be seen below that
$a \sim \alpha$ and thus $\lambda \sim 1$
near the phase transition point $g=g_c$ (we will see also that
$\Sigma \sim \alpha$ and thus it is much smaller than the $\omega^2$
term at high frequencies $\omega \gg \alpha^{1/2}$).
It means that the parameter
$a$ can be considered as a smooth function of $g$ in the vicinity
of $g_c$. Clearly, this conclusion does not depend on the model form
of $\tilde{G}_0(\omega)$ used in the above analysis.

 Now we re-introduce $\Sigma(\omega)$ into the equations for
$\hat{G}(\omega)$
and focus on its low-frequency behavior at $\omega \leq \sqrt{\alpha}$:
\begin{equation}\label{G_Q}
\hat G(\omega)=\frac{\alpha}{a - 2\Sigma(\omega) + \omega^2}\ ,
\end{equation}
\begin{equation}\label{Sigma1}
\Sigma(\omega) = \tilde{g}^6
\int\limits \hat G^3(t) \exp(i\omega t)\ dt
\end{equation}
where $\tilde{g} = g\chi_3^{1/3} \sim g$.
Strictly speaking, Eqs.(\ref{G_Q},\ref{Sigma1}) do not form closed system
since $a$ should be determined with the use of Eq.(\ref{init_cond}) which
contains high-frequency contributions. However, in this high-frequency
region (which produces the main contribution to the normalization
condition (\ref{init_cond}) ) the contribution of $\Sigma(\omega)$
can be neglected and thus
$a$ can be treated as an external control parameter which governs the
transition.
Green function defined by the Eqs.(\ref{G_Q},\ref{Sigma1}) acquires
singularity when $2\Sigma(0) = a$.  To find the form of this singularity,
we make use of the scaling Anzats $G(t) = qt^{-\nu}$ and
neglect $\omega^2$ term in the denumenator of Eq.(\ref{G_Q}). Then we
find $\nu =1/2$ and $q\sim \tilde{g}^{-1}\alpha^{1/4}$.
This critical-point solution
matches the short-time asymptotics (\ref{G0}) at $ t \sim \alpha^{-1/2}$.
The estimate for $\Sigma(0)$ which follows from the above scaling Anzats,
$$\Sigma(\omega = 0)\sim \tilde{g}^6 q^3\int\limits_{\sqrt a}^\infty
\frac{dt}{t^{3/2}} \sim \tilde{g}^3 q^3 a^{1/4}
$$
 gives
$\Sigma(0) \approx a$ at $\tilde{g}\sim 1$ and $a \sim \alpha$,
as it was expected.
These estimates show that second-order phase transition with critical
slowing down {\it may} indeed occur in the above range of parameters.
In the next section we will study the vicinity of the critical point
in more details.

\section{Green function near the $T=0$ transition point}

To study the form of the critical singularity, it is
convinient to define universal scaling functions ${\cal G}(\omega)$ and
$\sigma(\omega)$ which do not contain
small parameter $\alpha \ll 1$, and the parameter $b$ measuring the proximity
to the critical point:
\beq
\hat{G}(\omega) = {\cal G}(\tilde{\omega}); \quad
\alpha\Sigma (\omega) = \sigma(\tilde{\omega});
\quad \tilde{\omega} = \omega/\sqrt{\alpha}; \quad
b = (a-2\Sigma(0))/\alpha
\label{new}
\eeq
Now Eqs.(\ref{G_Q},\ref{Sigma1}) acquires the following form:
\begin{equation}\label{para}
{\cal G}(\tilde\omega)=\displaystyle\frac{1}{b + 2\left(\sigma(0) -
\sigma(\tilde\omega)\right)}; \qquad
\sigma(\tilde\omega) =
\tilde{g}^6 \int\limits {\cal G}^3(\tilde{t})
\exp(i\tilde\omega \tilde{t})\ d\tilde{t}
\end{equation}
Exactly at the critical point $b=0$ the solution of Eq.(\ref{para})
is
\beq
{\cal G}(\tilde\omega) =
 \left(\frac{\pi}{8}\right)^{1/4}\tilde{g}^{-3/2}|\tilde\omega|^{-1/2}
\label{critsol}
\eeq
Consider now the vicinity of the critical point, $0 < b \ll 1$.
It is clear form the form of the solution (\ref{critsol}) that
the similar result should be valid at $\tilde\omega \gg b^2$.
Next we focus on the long-time, low-$\omega$ region, $\tilde\omega \ll b^2$
and will look for the purely exponential solution
\beq
{\cal G}(\tilde{t}) = {\cal G}_1\exp(-\tilde{t}/\tau_0).
\label{expo}
\eeq
This type of asymptotic solution is known to exist in the classical
 version of the same model (cf.\cite{Chandra96a,Chandra97}). In the present
problem, one can show, considering analytic structure of Eqs.(\ref{para}),
that at $b > 0$ the singularity of ${\cal G}(\tilde\omega)$ which is closest
to the real axis of $\omega$, is necessarily a simple pole at some
$\tilde\omega = i/\tau_0$; the next singularity may exist at
$\tilde\omega \geq 3i/\tau_0$.
Solution of Eqs.(\ref{para}) with the Anzats (\ref{expo}) in
the region $\tilde{t} \gg \tau_0$
determines parameters $\tau_0$ and ${\cal G}_1$ as functions of $b$:
\beq
\tau_0 = \frac{(32/27)^{1/2}}{\tilde{g}^{3}}\frac{1}{ b^{2}} \quad
\quad {\cal G}_1 =(27/2)^{1/2} \tilde{g}^3\ b
\label{exposol}
\eeq
This solution is similar to the one found in~\cite{Chandra96a}; however, an
important difference is that in the present case the prefactor
${\cal G}_1$ scales to zero at the critical point $b=0$.

The full solution in the vicinity of the transition point should
contain both (\ref{critsol}) and (\ref{exposol}) as asymptotic solutions,
and can be written in the form
\begin{equation}
{\cal G}(\tilde{t}) =
\frac{1}{\sqrt{\tilde{t}}} f\left(\frac{\tilde{t}}{\tau_1}\right) +
{\cal G}_1\exp\left(-\frac{\tilde{t}}{\tau_0}\right)\ .
\label{fullsol}
\end{equation}
where $f(x)$ is some scaling function approaching constant at $x=0$ and
fast decaying at $x\to\infty$, and $\tau_1 \leq\tau_0/3$.
To confirm an existence of this type of solution, we solved Eqs.(\ref{para})
numerically for several values of $b \ll 1$.
The results of this computation are shown on Fig.~1. Clearly,
all three functions ${\cal G}(\omega)$ coincide in the high-$\omega$
region, there they are close the square-root asymptotic (\ref{critsol}).
Low-frequency parts (for $\omega \leq 0.08$) of these solutions
can be made coinciding
by a proper rescaling of their arguments,
$\omega^* = \Lambda\omega$. Fig.~2 demonstrates linear  relation
between $b^{-2}$ and the scaling coefficient $\Lambda$, as it was suggested
by Eqs.(\ref{exposol},\ref{fullsol}).
\begin{figure}
\epsfxsize=0.9\textwidth\epsfbox{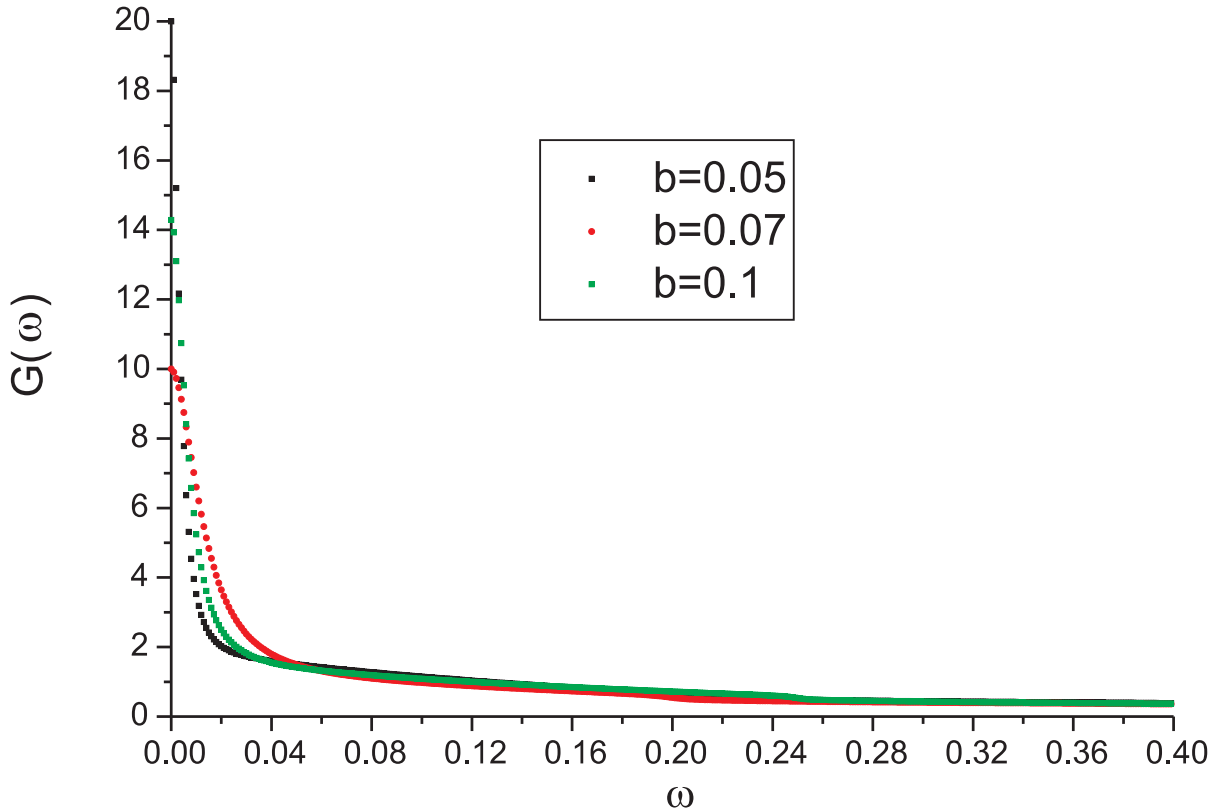}
\caption{Low-frequency asymptotic behavior of $\tilde{G}(\omega)$ at
different $b$ at $T=0$.}
\label{numerics1}
\end{figure}

\begin{figure}
\epsfxsize=0.9\textwidth\epsfbox{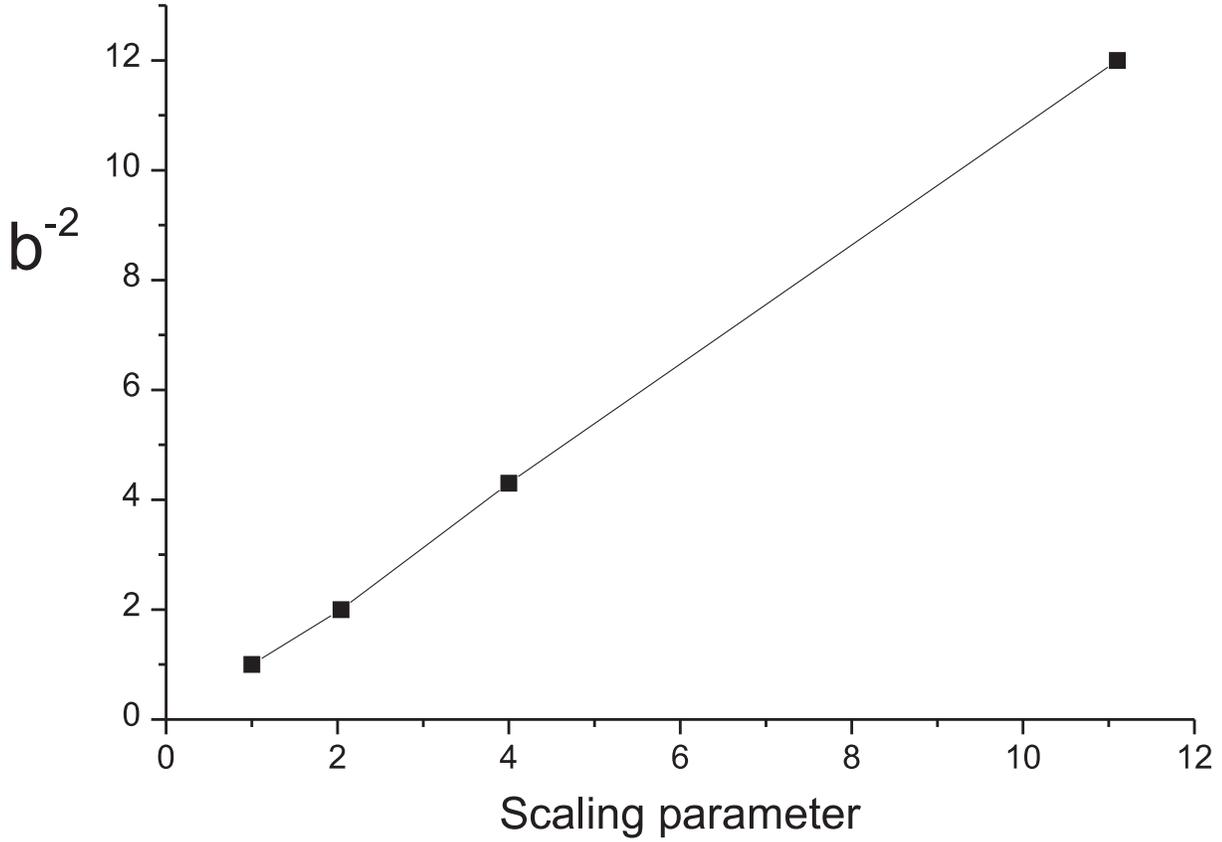}
\caption{The relation between scaling parameter $\Lambda$
and the proximity to the transition point $b$ at $T=0$.}
\label{numerics2}
\end{figure}
These results confirm the existence of the $T=0$ critical behavior
of the type of Eq.(\ref{fullsol}).

\section{Critical behavior at $T>0$}

The above results refer to the zero-$T$ phase transition controlled by
the single parameter $g = J_0/\sqrt{\alpha}$.
We found that this phase transition is a continuous one
and the corresponding critical behavior differs considerably from
the one found in an analogous classical model~\cite{Chandra96a}.
In particular, at $T=0$ critical point $g=g_c$
there is no ``plato'' solution with  approximately
constant $G(t)$ at $t \to \infty$ which is known to be a peculiar property
of regular classical glasses.
Now we consider low but non-zero temperatures $T=\beta^{-1}$ and will
find how  ``classical'' critical scaling ``grow up'' from the
``quantum'' background; we also find the low-temperature shape of the
phase transition line on the plane $(T,g)$.

Green function is defined
now at discrete frequencies $\omega_n=2\pi nT$ and
the equations~(\ref{G_Q}), (\ref{Sigma1}) can be written as
\begin{equation}\label{G_T}
 \hat G(\omega_n)=\frac{\alpha}{a -
2\Sigma(\omega_n) + \omega^2} , \qquad
\Sigma_M(\omega_n) =
\tilde{g}^6\int\limits_{0}^{\beta} \hat G^3(t) \exp(i\omega_n t)\ dt
\end{equation}
It will be convenient now to perform analytic continuation
of Eqs.(\ref{G_T}) and rewrite them in terms of real-time correlation
function $D(t) = \langle[S(t),S(0)]_+\rangle$ and response function
$\chi(t) = i [S(t),S(0)]_-\theta(t)$.
The functions $G(\omega_n)$, $D(\omega)$ and $\chi(\omega)$
are related as follows:
\beq
G(-i\omega+\eta) = \chi(\omega),\ \eta\to+0\ ,\qquad
D(\omega) = \mbox{Im}\chi(\omega) \coth(\omega/2T)
\label{Trelations}
\eeq
After analytic continuation the Eqs.(\ref{G_T}) can be written as
\begin{equation}\label{claschi}
\chi(\omega)=\frac{\alpha}{\tilde a - 2\Sigma(\omega)};\quad
\Sigma(\omega)= 8\tilde{g}^6
\int\limits^\infty_0 D^2(t)
\chi(t)\left( \exp(i\omega t) - 1 \right)\ dt;\quad
\tilde a = a - 2\Sigma(\omega = 0)
\end{equation}
where we omitted $\omega^2$ term which is irrelevant in the vicinity
of the critical point. Equations (\ref{claschi}) form (together with
the Fluctuation-Dissipation relation (second of Eqs. (\ref{Trelations}))
a closed set which determines critical singularity at $T >0$.
Formally Eqs.(\ref{claschi}) coincide with the corresponding ``classical''
equations from~\cite{Chandra96a}, the only difference is in the form of the
Fluctuation-Dissipation relation.

Let us consider low temperature region $T \ll \sqrt \alpha$.
As long as we are interested in the long-time asymptotic $ t \gg 1/T$,
the correlation and response functions
are related by classical FDT: $D(\omega) = 2T/\omega\ {\rm
Im}\,\chi(\omega)$.
Characteristic times, which is relevant
in~(\ref{claschi}), are also belong to classical region $t\gg 1/T$.
Therefore the correlation function at the transition point has the same critical
behavior as in the classical case: $\lim_{t\to\infty} D(t)=q$. However
parameter $a\equiv \lambda - g$ is determined by the "quantum" region of
 frequencies $\omega \gg T$, i.e. by the
equation~(\ref{eqfora}). The substitution of this expression to the
equations (\ref{claschi}) allows us to find
\begin{equation}
q \sim \alpha^{1/4} T^{1/2}; \qquad  \tilde{a} \sim \alpha^{3/4} T^{1/2}
\label{thermalq}
\end{equation}
In the short-time domain $t \ll T^{-1}$ the zero-$T$ critical
 solution with $D(t) \sim \alpha^{1/4}t^{-1/2}$ is valid.
Equation (\ref{thermalq}) demonstrates the way the "classical" solution
with nonzero $\lim_{t\to\infty} D(t)$ grows up with the temperature increase.

\section{Diamagnetic response near the transition point}

Correlation and response functions $D(t)$ and $\chi (t)$ are not directly
measurable in our system, but they can be used in order to calculate
measurable physical quantity which is dynamic diamagnetic susceptibility
$\chi_{\cal M}(\omega)$, like it was done previously for the classical
problem~\cite{Chandra96a}.
Total magnetic moment induced by time-dependent external magnetic field
is given by
\begin{equation}
{\cal M} = \frac{1}{2} \left(\frac{2e}{\hbar c}
\right) l^2
\sum_{mn} S_m^\dagger {\tilde{\cal J}}_{mn} S_n\ ,
\end{equation}
where ${\tilde{\cal J}}_{mn}= imn{\cal J}_{mn}$~\cite{Chandra96a}.
Then magnetic susceptibility $\chi_{\cal M}$ can be found making use of the
 Kubo formula:
$
\chi_{\cal M}(t-t') = i\left[ {\cal M}(t),\ {\cal M}^\dagger(t')\right]
\theta(t)
$
which leads to the expression
\begin{equation}
\chi_{\cal M}(\omega) = \left( \frac{2e}{\hbar c}\right)^2 l^2
\int_0^\infty \left(e^{i\omega t}-1\right) {\mbox Re}{\mbox Tr}
{\tilde{\cal J}} {\hat \chi(t)} {\tilde{\cal J}} {\hat D(t)} dt\ .
\label{chi_M}
\end{equation}
Here we omit the term, containing irreducible four-spin correlator (of
the order of $1/N$), and take into account that ${\cal M}(H=0) = 0$.
Note, that equation~(\ref{chi_M}) formally coincides with classical
formula for magnetic response~\cite{Chandra96a}. The matrix functions $\hat
D(t)$ and $\hat \chi(t)$ contain elements (denoted by superscript $^{(0)}$)
 belonging to the same (horizontal or vertical) sublattice of our array,
as well as ``offdiagonal'' elements (with superscript $^{(1)}$) which describe
correlation of phases on wires of different
type (horizontal/vertical). Relation between these functions is as follows:
${\hat \chi(\omega)}^{(1)} =J \widetilde G(\omega)
{\hat\chi(\omega)}^{(0)}$. Thus, the expression for magnetic susceptibility has
the following form:
\begin{equation}
\chi_{\cal M}(\omega)= \! \left( \frac{2e}{\hbar c} \right)^2 \!
\left(\frac{l^2} {12}\right)^2 N^5 \frac{J_0^2}{\alpha^2}
I(\omega)
\label{Idef}
\end{equation}
where
\begin{equation}
\label{Iomega}
I(\omega) = \int
\left(\delta(t-t_1)-\frac{J_0^2{\widetilde G}^2_(t-t_1)}{\alpha} \right)
\chi (t_1) D (t_1)\left(e^{i\omega t}-1\right)\theta (t)dtdt_1
\end{equation}

Near the transition point only long-time parts of all the function
in (\ref{Iomega}) are relevant, and this expression can be reduced to the
form
\begin{equation}
I(\omega) =
\left(\Sigma(\omega)-\Sigma(0)\right)
\int
\chi(t) D(t)e^{i\omega t}dt\ ,
\label{Iomega1}
\end{equation}
where the first factor came from the first brackets in (\ref{Iomega});
note that it vanishes in the limit $\omega\to 0$.

Using the solution (\ref{critsol}) we
obtain at the quantum critical point $J=J_c$:
\begin{equation}\label{chi_m_zero_high}
I(\omega) = \frac{\alpha}{2\pi}
\left(\frac{\alpha\pi}{8\tilde g^6}\right)^{1/4}
\sqrt{i\omega}\ln{\omega}
\end{equation}

Near the $T=0$ transition point at high enough frequencies
$\omega\gg (J/J_c - 1)^2\alpha^{-3/2}$ equation (\ref{chi_m_zero_high})
still holds.
In opposite case of low frequencies
\begin{equation}\label{chi_m_zero_low}
I(\omega) = \frac{8 J_c^3 \alpha^3}{81 (J_c-J)^3}
\frac{\omega^2\alpha^{1/2}}{g}\ .
\end{equation}
Note that the parameter $\tilde{g}$ (which is known up to the factors
of order 1 only) does not enter the low-$\omega$ asymptotic of $I(\omega)$.

Making use of the
Eqs.(\ref{Idef},\ref{chi_m_zero_high},\ref{chi_m_zero_low}) and returning
to the original units  of frequency, we obtain finally
ac diamagnetic susceptibility near the quantum transition point:
\begin{equation}
\chi_{\cal M}(\omega) \approx \left( \frac{2e}{\hbar c} \right)^2 \!
l^4 N^5 \frac{(J_cC_l)^{1/2}}{2e}
\sqrt{i\omega C_l/e^2}\ln(\omega C_l/e^2),\quad
\omega\gg \frac{C_l(J-J_c)^2}{e^2\alpha^{5/2}}
\end{equation}
\begin{equation}
\chi_{\cal M}(\omega) = \left( \frac{2e}{\hbar c}
\right)^2 \! \left(\frac{l^2}{12}\right)^2 N^5 \
\frac{2 C_l \alpha^{7/2}}{81 e^2 J_c}
\frac{J_c^3 }{(J_c-J)^3}\ \omega^2,\quad
\omega\ll \frac{C_l(J-J_c)^2}{e^2\alpha^{5/2}}\end{equation}
The above expressions are valid at the frequencies $\omega \gg T/\hbar$,
otherwise "classical" asymptotic for the Green functions should be
used and will lead to the frequency dependencies like those
in~\cite{Chandra96a}.

\section{Conclusions}

We have shown that  regularly frustrated long-range Josepshon array has a quantum
(zero-temperature) phase transition between Coulomb-dominated insulator phase and a
superconductive state. This transition happens when the nearest-neighbors Josephson
coupling exceeds the critical value, $J_{ij} \sim N^{-1/2}\sqrt{\alpha} e^2/C_l$, where
$C_l$ is the self-capacitance of an individual wire.

We found that quantum critical behavior of the model at $J \to J_c$ is different
from that of an analogous classical system~\cite{Chandra96a}: at the quantum critical point
$D(t) \sim t^{-1/2}$ while at the classical critical point $q =\lim_{t\to\infty} D(t)$.
However, at any non-zero temperature a "classical" type of asymptotic behavior is recovered
at the longest-times, $t \gg \hbar/T$, leading to $q\propto T^{1/2}$.
Near the $T=0$ critical point the gap in the excitation spectrum decreases
as $\tau_0^{-1} \propto (J_c-J)^2$.

Near the phase transition the effective inductance ${\cal L}$ of the array
(defined by ${\cal L} \propto
\partial^2\chi_{\cal M}(\omega)/\partial\omega^2|_{\omega\to 0}$)
diverges  as $(J_c-J)^{-3}$, this shows that the glass state
has a macroscopic phase rigidity (cf. also~\cite{Vinokur87}).
Right at the critical point we find unusual frequency behavior of the
complex diamagnetic susceptibility,
$\chi_{\cal M}(\omega) \propto \sqrt{i\omega}\ln\omega$.

Frustrated nature of couplings in our array and comparison with the
previous results~\cite{Chandra97} on the classical version of the same model indicates
that the high-$J$ state is a {\it quantum glassy superconductor}.
The $T=0$ nonergodic properties (irreversibility, ageing) remain an open question;
note here that recent study~\cite{Cu98} of nonequilibrium glassy behavior in a p-spin
spherical quantum model assumed strongly dissipative (overdamped) dynamics of the
whereas dynamics relevant for the Josephson array at $T=0$ must be underdamped.

\end{document}